\documentclass{elsart}

\usepackage{amssymb}
\usepackage{epsfig}
\begin{document}
\begin{frontmatter}

\title{Electrical Properties of Carbon Fiber Support Systems }


\author[Fermilab]{W.~Cooper,}
\author[UW]{C.~Daly,}
\author[Fermilab]{M.~Demarteau,\corauthref{cor}}
\ead{demarteau@fnal.gov}
\author[Fermilab]{J.~Fast,}
\author[Fermilab]{K.~Hanagaki,}
\author[Fermilab]{M.~Johnson,\corauthref{cor}}
\ead{mjohnson@fnal.gov}
\author[UW]{W.~Kuykendall,}
\author[UW]{H.~Lubatti,}
\author[Fermilab]{M.~Matulik,}
\author[Fermilab]{A. Nomerotski,} 
\author[olemis]{B.~Quinn,}
\author[UW]{J.~Wang}

\address[Fermilab]{Fermi National Accelerator Laboratory, Batavia, IL 60510}
\address[olemis]{University of Mississippi, Department of Physics, 
                 University, MS 38677}
\address[UW]{University of Washington, Department of Physics, 
             Seattle, WA 98195}
\corauth[cor]{Tel.: 630-840-2840 (Demarteau), 630-840-3168 (Johnson)}

\begin{abstract}
Carbon fiber support structures have become common elements 
of detector designs for high energy physics experiments. 
Carbon fiber has many mechanical advantages but it is also 
characterized by high conductivity, particularly at high frequency,
with associated design issues. This paper discusses the 
elements required for sound electrical performance of 
silicon detectors employing carbon fiber support elements. 
Tests on carbon fiber structures are presented indicating 
that carbon fiber must be regarded as a conductor for the 
frequency region of 10 to 100~MHz. 
The general principles of grounding configurations involving 
carbon fiber structures will be discussed. 
To illustrate the design requirements, 
measurements performed with a silicon detector on a carbon fiber 
support structure at small radius are presented. 
A grounding scheme employing copper-kapton mesh circuits is described
and shown to provide adequate and robust detector performance. 
\end{abstract}

\end{frontmatter}

\section{Introduction}
\label{intro}
Nowadays carbon fiber is a ubiquitous structural material 
used in a wide range of applications due to 
its high modulus and low mass.  
It offers great flexibility in terms of tuning thermal 
and mechanical properties through the orientation and number of 
lay-ups of the fibers. Carbon fiber support structures 
have also become common elements of detector designs for high 
energy physics experiments and are especially prevalent 
in the design of silicon vertex detectors. 
Early examples of the use of carbon fiber in external 
support structures are the DELPHI silicon detector~\cite{ref:delphi}, 
and the space frames for the BaBar~\cite{ref:babar} and 
CDF ISL silicon detectors~\cite{ref:isl}. 
In current designs, carbon fiber 
is integrated in the design of silicon detector modules, 
like the readout modules for the silicon tracker 
for the CMS experiment~\cite{ref:cms}. 
An example of a detector with a 
fully integrated carbon fiber support structure  
is the Layer~00 detector for the CDF collaboration at the 
Fermilab Tevatron collider~\cite{ref:layer00}. 

While the use of carbon fiber solves a variety of mechanical 
problems, it presents a challenging set of electrical concerns.  
Highly conductive carbon fiber surfaces produce an undesirable
capacitance relative to sensors, electronics and cables and 
can compromise the detector performance 
due to significant coherent noise pickup~\cite{ref:l00_noise}. 
Well-designed coupling and grounding schemes are 
essential for producing low-noise environments for these 
detectors.  

Here we describe the electrical characteristics of carbon fiber 
as applicable to a silicon tracker at small radius with an 
integrated carbon fiber support structure. 
Measurements are performed on a prototype of a detector 
that has been proposed for the D\O\ experiment 
at the Fermilab Tevatron collider, called the Layer Zero 
(L\O) detector~\cite{ref:TDR}. 
In the next section the electrical characteristics of carbon 
fiber will be described. The carbon fiber test pieces were made
using carbon fiber epoxy resin prepregs using 
K139, obtained from Bryte~\cite{ref:bryte}, 
and K13C, obtained from YLA~\cite{ref:yla}. 
The carbon fiber in both products was manufactured by Mitsubishi. 
K139 uses 110 Msi modulus fiber with a thermal conductivity of 
210~W/mK, with a fiber aerial weight (FAW) of 55~g/m$^2$. 
K13C has a modulus of 130 Msi, with a significantly higher thermal 
conductivity of 620~W/mK and a FAW of 69~g/m$^2$~\cite{ref:asm}. 
For both prepregs, the resin fraction is about 35\% 
by weight, or roughly 50\% by volume before curing.  

Based on the results of the electrical tests of the carbon fiber, 
methods will be outlined that address the 
electrical requirements for the construction of support structures 
for silicon detectors. 
The design addresses all potential noise sources. 
Tests aimed at minimizing coherent noise contributions with a 
full-scale silicon detector, populated with a 
limited number of readout modules will be described. 
The paper concludes with a suggested set of design rules
which need to be respected in the construction of silicon 
detectors with integrated carbon fiber supports.

\section{Electrical Conductivity of Carbon Fiber}
\label{carbon}

Carbon fiber with ultra-high modulus ($\sim$1000 GPa) also has 
low resistivity (100~$\Omega$cm)~\cite{ref:asm}.  
These materials are characterized by particularly high 
conductivity at high frequency with associated design issues. 
A series of tests were carried out to verify the conductivity 
of carbon fiber. 

The first study performed was a measurement of capacitor 
impedance. Two parallel plate capacitors were built using 
$6'' \times 6''$  bare FR4 cores, with $\epsilon_r = 4$, 
approximately $0.27''$ thick. 
The first capacitor, used for baseline measurements, had two 
tinned copper electrodes laminated to either side of the 
FR4 dielectric. Each electrode is approximately $0.075''$ thick.
The second capacitor has a single tinned copper electrode 
attached to one side of the FR4 dielectric and a 
single carbon fiber electrode attached to the other side. 
The carbon fiber electrode was four plies of K139 fiber with 
a 4-layer $[0^{0}/90^{0}]_{s}$ lay-up, approximately $0.04''$ thick.
Contact with the two electrodes of the capacitor under measurement 
was established using two 20AWG stranded wires connected to the 
two contacts of a BNC connector. For the capacitor electrode side, 
the strands were separated and splayed to maximize contact and 
taped to the electrode with a specific area of copper tape.  
Prior to the attachment of the probe, the contact area on the 
electrode was burnished with mildly abrasive polishing material 
(Scotch-Brite) and cleaned with ethyl alcohol.  After taping 
the probe to the electrode, the tape was worked with a blunt tool 
to maximize contact to the strands of the probe and to force air 
from under the copper tape to the outside edge to maximize contact 
to the electrode.  Past experience has shown that not taking care 
at this stage in the test preparation would result in inconsistent
measurements. The size of the copper tape used to make contact with 
the carbon fiber electrode was varied to measure the effect of the 
fraction of contact area required to obtain a given impedance 
response.

\begin{figure}[h]
\begin{center}
\resizebox{!}{3.0in}{\includegraphics{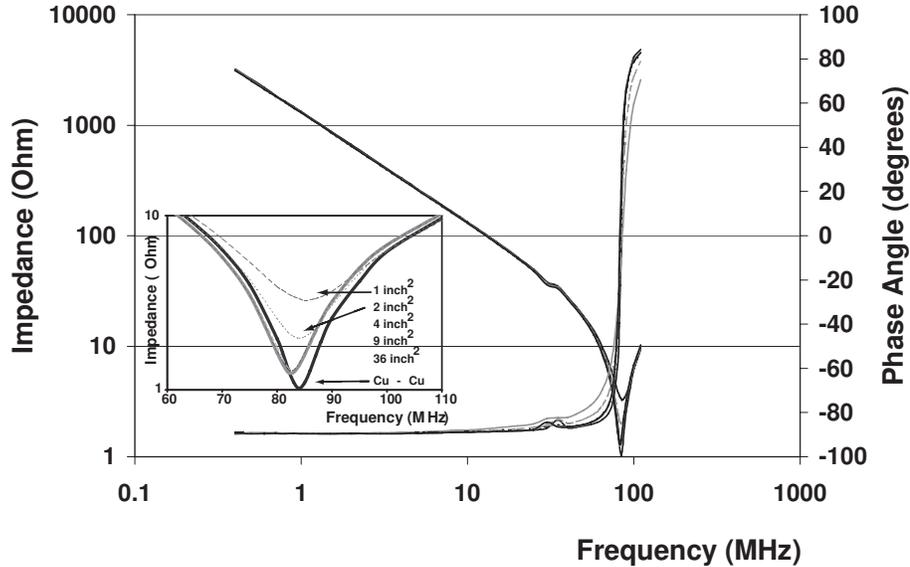}}
\caption{Impedance and phase shift as function of frequency 
for different configurations of parallel plate capacitors. }
\label{fig:imp}
\end{center}
\end{figure}

Impedance measurements were then made using an HP4193A Vector
Impedance meter at specific frequencies over a range of 
0.4 to 110~MHz. The measurements were performed 
in a temperature and humidity controlled environment.
The measured impedance and phase shift is shown 
in Fig.~\ref{fig:imp} for six different capacitors: the reference 
capacitor with two copper electrodes, and the capacitors with one 
copper electrode and one carbon fiber electrode with a copper tape 
contact area of 1, 2, 4, 9 and 36~square inches. The inset shows a 
blow-up of the region around the minimum. 
For low frequencies, the behavior is the expected $1/j \omega C$ 
dependence. In this frequency range, from 0.4 to 50~MHz, 
the response of all capacitor configurations is 
virtually identical. 
With increasing frequency the impedance decreases rapidly 
and reaches a minimum with vanishing phase shift. 
At this point the impedance is real and purely resistive. 
The reference capacitor with two copper electrodes reaches 
the lowest impedance of all configurations. The capacitor 
configurations with one carbon fiber electrode with a copper 
area of 4, 9 and 36~${\rm in}^2$ are virtually indistinguishable.

\begin{figure}[h]
\begin{center}
\resizebox{!}{3.0in}{\includegraphics{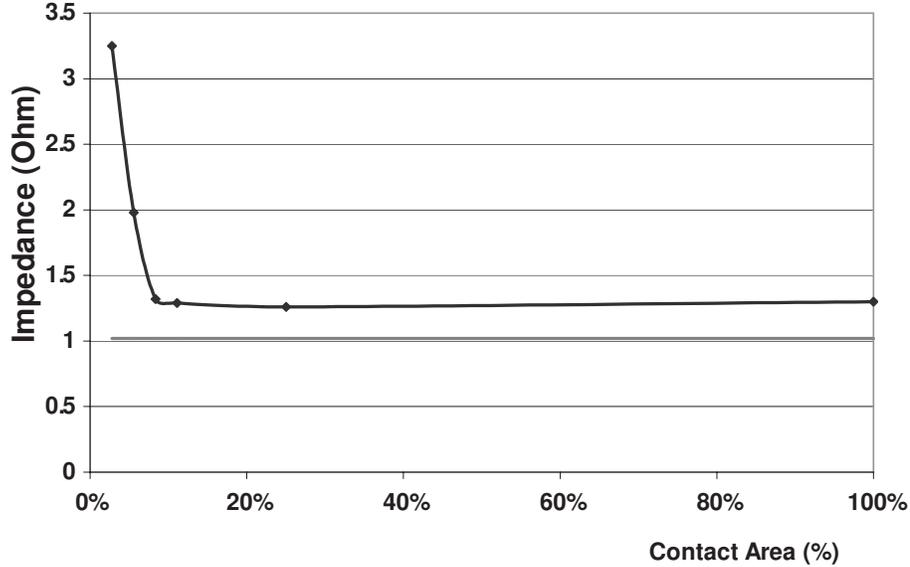}}
\caption{Impedance of parallel plate capacitors as function of 
         the copper area on the second, carbon fiber, electrode 
         (see text for details). }
\label{fig:imp_area}
\end{center}
\end{figure}

Plotting the value of the minimum of the impedance from 
Fig.~\ref{fig:imp} as function of fractional copper contact area 
of the carbon fiber gives the result shown in Fig.~\ref{fig:imp_area}. 
The frequency at which the impedance minimum occurred was at 
84$\pm$1~MHz for all capacitor configurations tested.
The impedance of the reference capacitor, with dual copper 
electrodes, is 1.02~$\Omega$ and is indicated by the solid 
line. 
The results show that at high frequencies, the frequencies of 
interest for silicon vertex detectors, carbon fiber is to be 
regarded as a conductor. Furthermore, a 15-20\% area coverage of 
the carbon fiber with a good conductor should provide adequate 
electrical coupling to the carbon fiber. 

The measurements were also performed by placing the capacitors 
in a bandstop circuit and measuring the power transfer 
function with an HP3577A network analyzer as function of frequency. 
The results are in complete agreement with the results quoted above. 
 
\begin{figure}[h]
\begin{center}
\resizebox{!}{5.0in}{\includegraphics{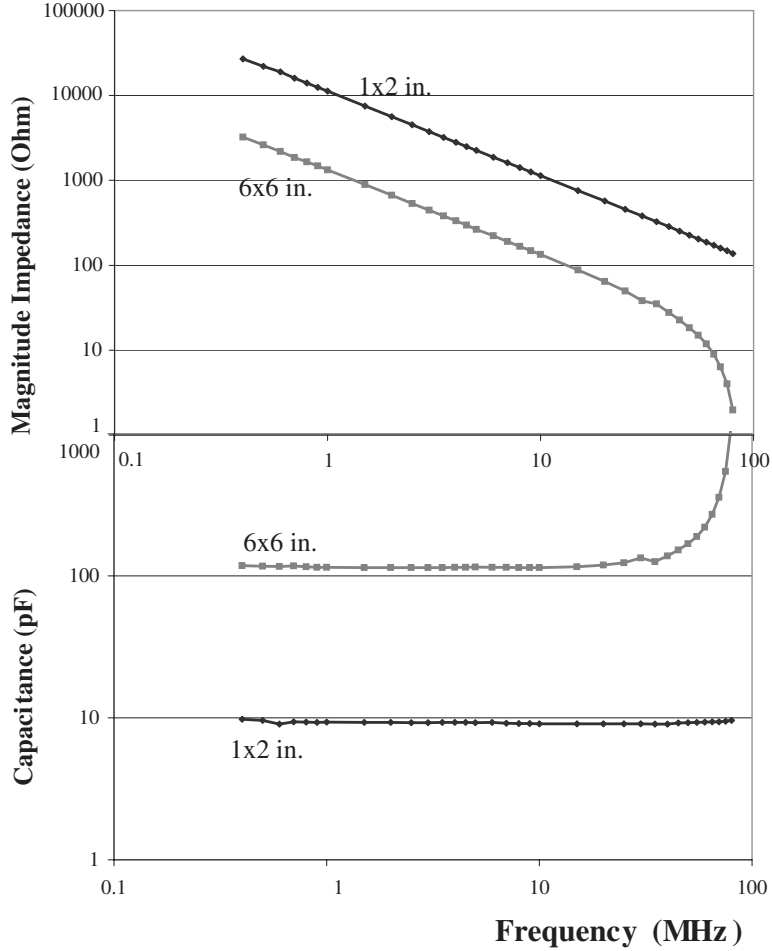}}
\caption{Impedance and capacitance as function of frequency 
for two different configurations of parallel plate capacitors. 
The top curve in the upper and lower plot corresponds 
to the capacitor with the carbon fiber plate removed and 
replaced with a 2~${\rm in}^2$ piece of copper directly attached 
to the dielectric; the other curve in either plot corresponds to the
capacitor with two copper electrodes. }
\label{fig:imp_nocf}
\end{center}
\end{figure}

To verify that the carbon fiber was not acting as part of the 
dielectric, another measurement was taken with the carbon fiber 
completely removed and with the 
1 by 2 inches copper tape directly attached to the FR4 sheet. 
The frequency response of the impedance for this configuration, 
together with the response of the reference capacitor with two 
copper electrodes, is shown in the upper half of
Fig.~\ref{fig:imp_nocf}. 
The upper impedance curve corresponds 
to the capacitor with the carbon fiber removed and replaced with 
copper tape with an area of 2~${\rm in}^2$. The difference 
between the measured impedance for the two configurations, 
compared to the frequency response with the carbon fiber sheet 
(see Fig.~\ref{fig:imp}) is striking and clearly shows that the 
carbon fiber does not act as part of the dielectric. 
The lower half of Fig.~\ref{fig:imp_nocf} shows the capacitance for 
the two configurations as function of frequency, corrected for the 
contribution from the test setup. Given an area 
ratio of 18 for the two configurations (36~${\rm in}^2$ versus 
2~${\rm in}^2$) the ratio of the two capacitance values is naively 
expected to be a factor of 18. A ratio of 11.9 is measured,
however. 

To understand the measurement, a full ANSYS calculation was 
performed~\cite{ref:ansys}. 
Using ANSYS, the capacitance of 
the 6 by 6 inch capacitor with the copper electrodes was calculated 
to solve for the dielectric constant of FR4. Based on the measured 
value of $117 \pm 6$~pF, the ANSYS calculation gives a value for the 
dielectric constant of $3.6 \pm 0.2$. Using the simple analytic 
formula for the capacitance of area over separation, 
a value of 3.9 is obtained. 
ANSYS predicts 
a capacitance of 8.0~pF for the capacitor with the 2~${\rm in}^2$ 
copper electrode, whereas the measured value, corrected for 
capacitance of the leads, is $9.8 \pm 1.5$~pF. The contributions 
to the error estimate are a 10\% error on the measurement itself, 
a 10\% error on the correction due to the test fixture, 
and a 5\% error on the extraction of the dielectric constant, 
added in quadrature. 
The measured ratio of capacitances of 11.9~$\pm$~1.9 is in 
reasonable agreement with the predicted ratio of 14.6,  
validating our conclusion that carbon fiber does not act as 
part of the dielectric.

These studies show that carbon fiber should
be regarded as a conductor for the frequencies above one MHz. 
Using highly conductive carbon fiber as a mechanical support
thus has the potential to exhibit strong capacitive coupling to 
the silicon sensors it supports, which can lead to very troublesome 
sources of noise, especially in close-packed structures. 
It is therefore imperative that all the carbon fiber in the 
detector is effectively shorted to the hybrid or bias filter 
grounds to prevent noise transmission through capacitive coupling 
to the sensor readout. 
It has been demonstrated that copper in contact with 
the carbon fiber with more than 10-15\% area coverage provides 
adequate electrical coupling.  In the next section methods 
to apply our findings to the construction of silicon detector
modules will be explored.

\section{Development of Grounding Method}
\label{method}


In the design of the silicon detector it is envisioned that the 
sensors will be mounted directly on the support structure and 
that a low-pass filter card for the bias high voltage and bias 
return is glued directly onto the sensor. The ground plane of 
the sensor and the carbon fiber support structure then form the 
plates of a parallel plate capacitor. 
This puts the sensors between two plates of a capacitor, 
and any power transfer through the capacitor caused by unequal 
potential at its plates will generate noise on 
the sensor.  The challenge is to devise a low mass short between the 
carbon fiber and ground plane which is highly efficient over a 
frequency range of 5 kHz - 50 MHz. 
To determine the feasibility of shorting the carbon fiber 
support cylinder, on which the  silicon sensors are 
mounted, to the hybrid or filter ground plane, a series of tests 
were carried out with a mockup of the real design.

\begin{figure}[ht]
\begin{center}
\resizebox{!}{2.5in}{\includegraphics{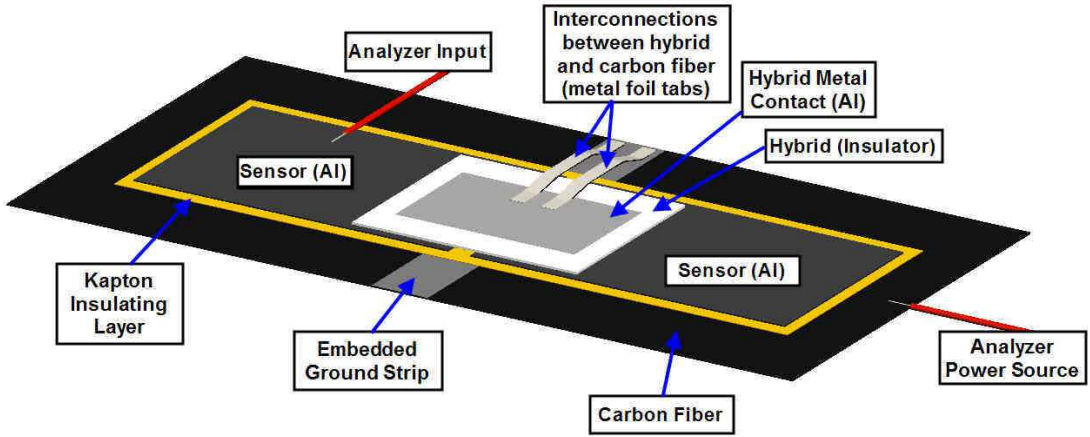}}
\caption{Sketch of the mockup of the detector for the coupling 
         study. Silicon sensors and the ground plane on a bias voltage 
         filter card or hybrid are modeled using 0.5~mil thick aluminum 
         strips, separated by a non-conducting plastic. 
         This sketch shows the configuration of one embedded ground
         strip on 
         the carbon fiber support. Up to four strips of three different 
         materials were used. }
\label{fig:mockup}
\end{center}
\end{figure}

A mock-up of the carbon fiber support/sensor/ground plane 
configuration was built (see Fig.~\ref{fig:mockup}). 
Two pieces of 0.5~mil aluminum 
(indicated by `Sensor (Al)' in Fig.~\ref{fig:mockup})
representing two sensors, 
were attached to a strip of kapton, that was mounted directly 
onto a carbon fiber support structure. 
This carbon fiber support piece has a 6-layer 
$[0^{0}/+20^{0}/-20^{0}]_{s}$ 
lay-up of K13C high modulus carbon fiber. 
It should be noted that K13C is even more conductive than K139.
A piece of plastic, mimicking the filter card or hybrid, 
was laid on top of the `sensors'. 
A thin 0.5~mil aluminum foil was then attached to the plastic 
representing the ground plane
(indicated by `Hybrid (Al)' in Fig.~\ref{fig:mockup}). 
The carbon fiber support was driven with the source power from 
the network analyzer. The network analyzer input was wired to 
the sensor aluminum layer and both the source and input grounds 
were connected to the filter plane
(`Hybrid (Al)' in Fig.~\ref{fig:mockup}). 
Power transfer functions 
were measured with different configurations for shorting the 
carbon fiber to the filter plane. 
One to four ${1 \over 8}''$ wide strips of 1~mil thick 
aluminized Mylar, 0.5~mil aluminum, or 1~mil copper grounding strips 
were attached between the kapton and the carbon fiber. 
The aluminized Mylar had a 250-300~\AA \ thick layer of pure aluminum. 
Short tabs of the same material were attached to the grounding 
strips and these strips were wrapped around the 
`sensors' and attached to the filter/hybrid  
ground plane (see Fig.~\ref{fig:mockup}).  

First, a reference power transfer function was measured with no 
grounding strips attached. Next, the grounding strips were attached
one by one, varying the area of the carbon fiber covered with grounding 
strips, and the power transfer was measured again. 
It was found that 
aluminized Mylar strips were 
minimally effective in reducing power to the sensor. 
Aluminum and copper strips performed equally well. 
The power reduction was independent of the number and size 
of strips, but proportional to 
the area used to couple the strips to the carbon fiber. 
We achieved  a maximum reduction in excess of 40~dB at 1~MHz with 
copper tape covering about 11\% of the carbon fiber, identical to the 
result shown in the previous section (see Fig.~\ref{fig:imp_area}). 

Because copper tape coupling would not be feasible in the detector
due to mass constraints, 
the next coupling test was carried out using carbon fiber pieces with 
embedded aluminum foil and aluminized Mylar.  
The aluminum or Mylar was attached to the carbon fiber 
by co-curing it with the carbon fiber.  
A new sensor/hybrid mockup was built and mounted on different pieces 
of K13C carbon fiber which had various amounts of surface area 
covered with embedded aluminum or aluminized Mylar. 
The embedded grounding material, as before, extends 
beyond the sensors so that grounding tabs can be attached to it 
and to the 
hybrid/filter plane, as sketched in Fig.~\ref{fig:mockup}.   
It was noticed that aluminized Mylar grounding produced 
significantly less attenuation 
than aluminum; about 10 dB less below 20 MHz.  
This test also verified that the attenuation is not affected by 
the width of the grounding strips.  
Measurements of transfer functions were taken with different 
areas of embedded aluminum and equal sized grounding strips.  
Those data confirmed that the attenuation increases with contact 
area to the carbon fiber.  The maximum area of embedded aluminum 
tested was about 4~in$^2$ (again, about 11\% 
of the area), giving a power decrease of at least 30~dB for 
all frequencies below 50~MHz. 

Although coupling to carbon fiber through 
embedded aluminum has thus been shown to be an effective grounding 
technique, the use of aluminum does not represent a robust solution. 
The aluminum is very susceptible to oxidation and the likely choice 
for contacts to the ground plane would be silver epoxy, which 
historically has not been particularly reliable. 
This has led us to the design which employs a 
kapton flex circuit with a copper mesh, embedded in the carbon 
fiber surface. To obtain a more robust design we opted for 
a configuration where the kapton covers the surface area 
completely, with the embedded copper mesh having only a limited 
area coverage.

\section{Full Prototype Experimental Setup}
\label{sec:setup}

The design of the L\O\ detector calls for the silicon sensors to 
be mounted directly on a full length carbon fiber support 
structure at a radius of about 17~mm. 
Sensors with $p$-implants on an $n$-type substrate with 
50~$\mu m$ readout pitch, 8~cm long, produced by ELMA~\cite{ref:elma}, 
were used in the study. The sensors have 256 readout strips. 
A low-pass filter card of 150~$\mu$m thick G-10 for the 
bias high voltage and bias return is glued directly  
onto the sensor, close to the wire bonding pads. 

Often the hybrid with the front-end readout electronics is 
mounted directly on the sensor. In this design, 
due to space constraints and heat dissipation, the signals from 
the silicon sensors are transmitted to the hybrid containing the 
readout chips by flexible circuits up to 435~mm long.  
Since these circuits carry analogue
signals they are referred to as `analogue cables'. 
Although this solution is very attractive because of reduced 
material and heat generation in the sensitive volume, it 
represents a considerable technical challenge. 
The addition of the analogue cable reduces 
the signal to noise of the silicon sensors due to the added 
capacitive load. 

To minimize the capacitance between the traces of the 
analogue cable, and thus optimize the S/N ratio, our 
design uses a stack of two cables 
with constant 91~$\mu$m pitch, laminated with a lateral 
shift of 45~$\mu$m. The pitch of the cable stack thus matches the 
pitch of both the readout strips on the sensor and of the
input bond pads on the front-end readout chip. 
It should be noted that with this configuration adjacent readout 
channels alternate between the top and bottom cable in the stack. 
Stated in a different way, all odd readout channels are connected 
to one cable in the stack and all even channels to the other cable. 
For our configuration, the first channel is channel~1, and all 
odd channels (1, 3, 5, $\dots$) correspond 
to the top cable in the stack. 
In order to reduce the capacitance contribution from the adjacent 
cable within the stack of two cables, a spacer consisting of 
150~$\mu$m thick polypropylene mesh with about 25\% 
volume occupancy is placed between the two cables. 
After a series of prototypes, flawless cables were produced 
by Dyconex~\cite{ref:Dyconex}. The capacitance as measured for 
this arrangement is 0.36~pF/cm in good agreement with 
ANSYS calculations. In our setup we used 42~cm long 
prototype cables. Each cable has 129 signal traces with  
91~$\mu$m pitch. Each trace is 16~$\mu$m wide copper on 
a 50~$\mu$m thick kapton substrate. One trace is reserved as 
a spare trace in case of an open. 
Two additional, wider traces (100~$\mu$m) are provided on the 
cable for the bias voltage and its return.

The analogue cable is wirebonded to a ceramic hybrid containing 
two SVX4 readout chips~\cite{ref:svx4} with 128 input channels 
each. 
The chip operates at 53 MHz.
The equivalent noise charge (ENC) of the SVX4 chip is 
$300 + 41*C({\rm pF})~(e^-)$, at a fixed risetime of 69~ns.   

\begin{figure}[htbp]
  \begin{center}
  \resizebox{!}{1.0in}{\includegraphics{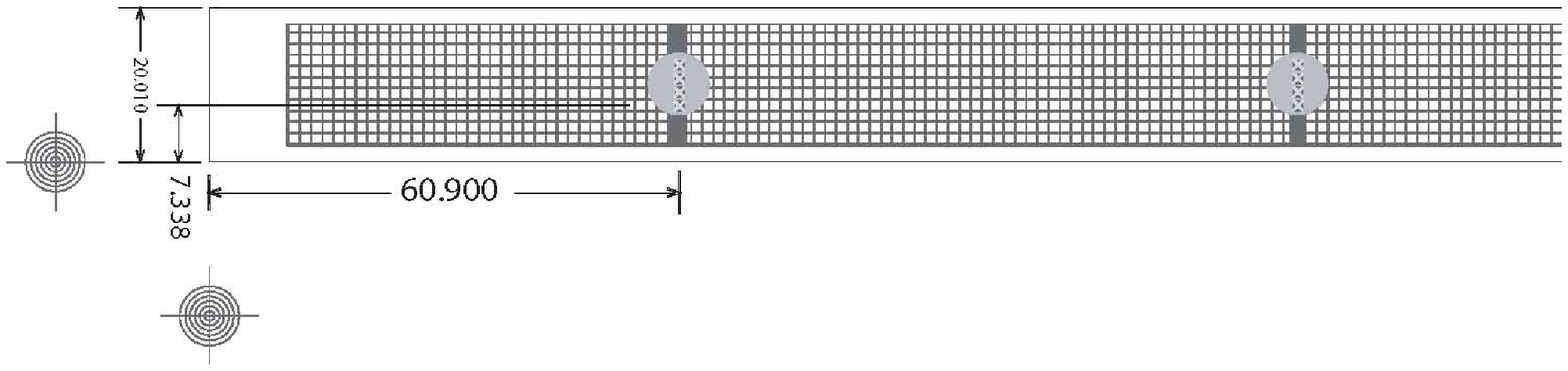}}
  \resizebox{!}{0.4in}{\includegraphics{null.eps}}
  \resizebox{!}{2.0in}{\includegraphics{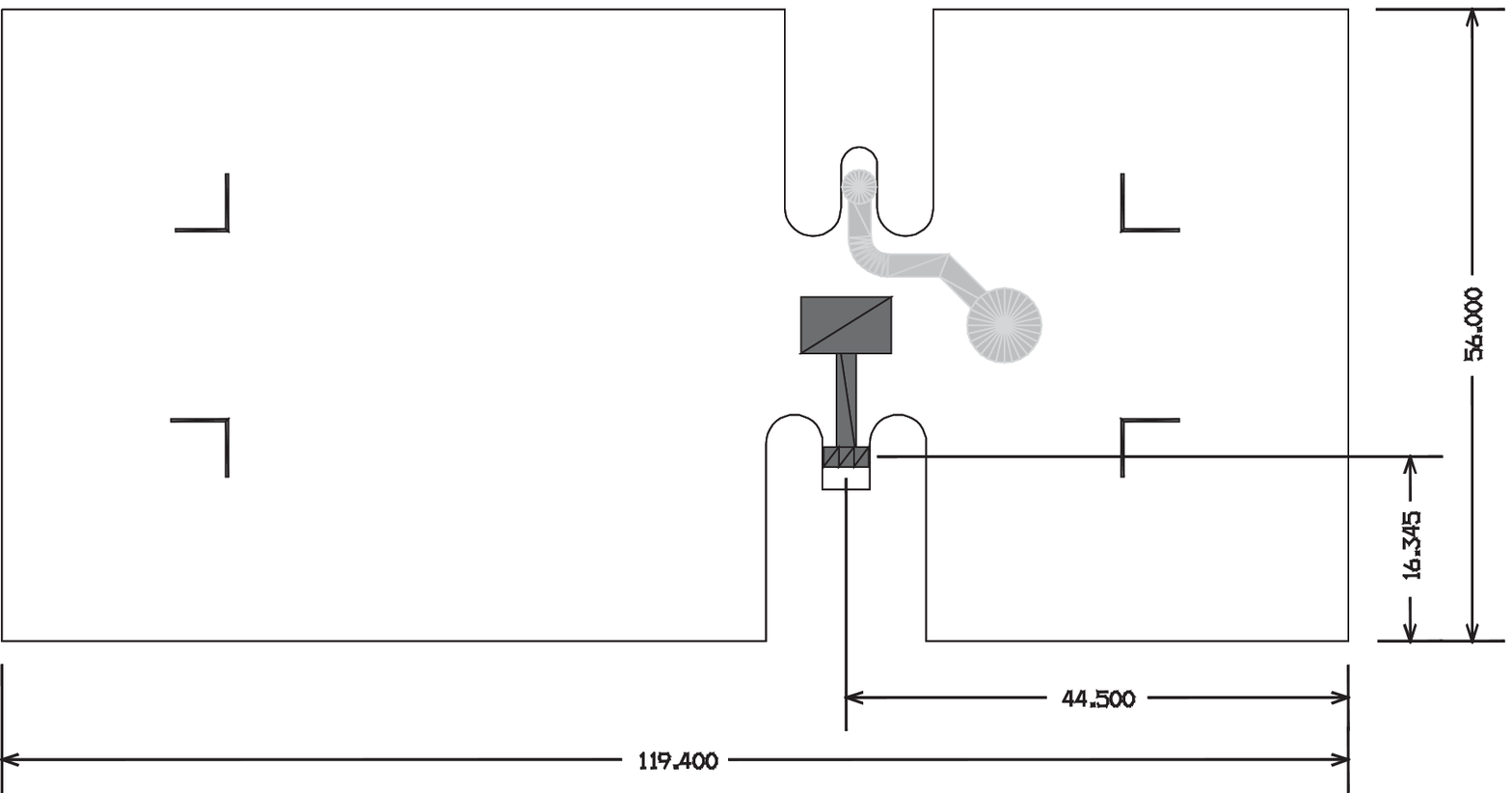}}
  \caption{Drawing of the copper mesh cable that was embedded in the 
           carbon fiber support structure (top); 
           drawing of the sensor `wrap-around' flex circuit (bottom).
          }
  \label{fig:cable}
  \end{center}
\end{figure}

\begin{figure}[h]
\centering
\resizebox{!}{3.5in}{\includegraphics{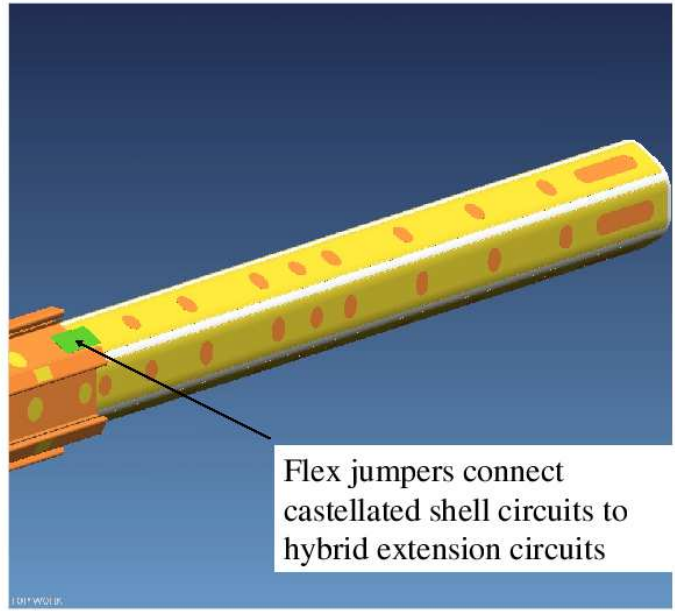}}
\caption{Solid model of the transition region between sensor and 
         hybrid support. The separate embedded ground mesh circuits 
         are visible. Also indicated is the jumper cable to 
         electrically connect the two separate ground mesh circuits. }
\label{fig:solidmodel}
\end{figure}

A full-size prototype crenellated two-part support structure 
was made using K13C fiber. It consists of a twelve-sided 
inner shell and a six-sided outer shell.  
The inner shell has a 4-layer $[0^{0}/90^{0}]_{s}$ 
lay-up of K13C high modulus carbon fiber.   
The outer shell has a 6-layer $[0^{0}/+20^{0}/-20^{0}]_{s}$ 
lay-up of the same high modulus fiber. The inner shell extends 
beyond the length of the outer shell and forms the support 
structure for the hybrids. The outer shell only supports the 
sensors (see Fig.~\ref{fig:solidmodel}). 
A 0.025~mm kapton sheet with an embedded 
copper mesh was co-bonded to the outer shell for grounding 
connections. 
The copper mesh, 5~$\mu$m thick, which is in direct contact with the 
carbon fiber, has an area coverage of 30\%. 
Round copper pads, one for each sensor, 
are provided on the top side of the mesh. The pads are nickel-gold 
plated, with a thickness of 1.3~$\mu m$ and 0.3~$\mu m$, respectively. 
Four vias, with 0.889~mm diameter holes establish the electrical 
connection between the pad and the copper mesh. Four vias were 
chosen to lower the inductance and for redundancy. 
The top drawing in Fig.~\ref{fig:cable} shows one end of the 
copper-mesh cable, produced by Compunetics~\cite{ref:comp}. 

\begin{figure}[h]
\centering
\resizebox{!}{2.0in}{\includegraphics{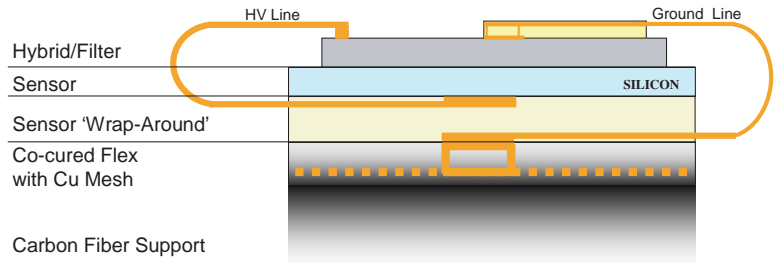}}
\caption{Schematic cross section of the electrical configuration 
         (drawing not to scale).   }
\label{fig:flex}
\end{figure}

An insulating two-layer kapton flex circuit, 
shown in the lower drawing in Fig.~\ref{fig:cable}, 
is glued to the backside of the sensors, covering the backside 
completely. This circuit, also produced by Compunetics, 
has one tab on either side that 
wraps around the sensor and is attached to the filter card mounted 
on the sensor. The right angles in the drawing indicate where the 
kapton is trimmed, so that the full backside of the sensor
is covered by kapton. 
After the backside lamination, the sensor is mounted 
on the support structure with the embedded mesh. 
The tab connected to the round pad in the bottom drawing in 
Fig.~\ref{fig:cable} picks up the bias voltage from 
the filter card and directly applies it to the backside of the sensor. 
The small rectangular pad connected to the larger rectangular pad 
is the ground connection. 
The smaller rectangular pad has three 0.889~mm vias which establish the 
connection to the co-bonded copper mesh of the 
outer shell of the support structure. 
Fig.~\ref{fig:flex} shows a schematic cross section of 
a sensor mounted on the support structure to aid in the 
understanding how all electrical connections are established.

The twelve-sided inner shell of the support structure is longer 
than the crenellated outer shell and forms the support structure 
for the hybrids. A separate kapton-copper mesh, with the same 
specifications as the circuit for the sensor support, was embedded 
in the carbon fiber to provide the ground for the hybrids. 
At the transition between the  
sensor and hybrid support regions the two copper mesh circuits 
can be connected electrically through a small jumper mesh 
cable. It should be noted that in our 
initial tests, described below, the two ground mesh circuits 
were not connected. Fig.~\ref{fig:solidmodel} shows a detail
of the solid model of the transition region between sensor and 
hybrid support regions.

\begin{figure}[htbp]
  \begin{center}
  \resizebox{!}{3.2in}{\includegraphics{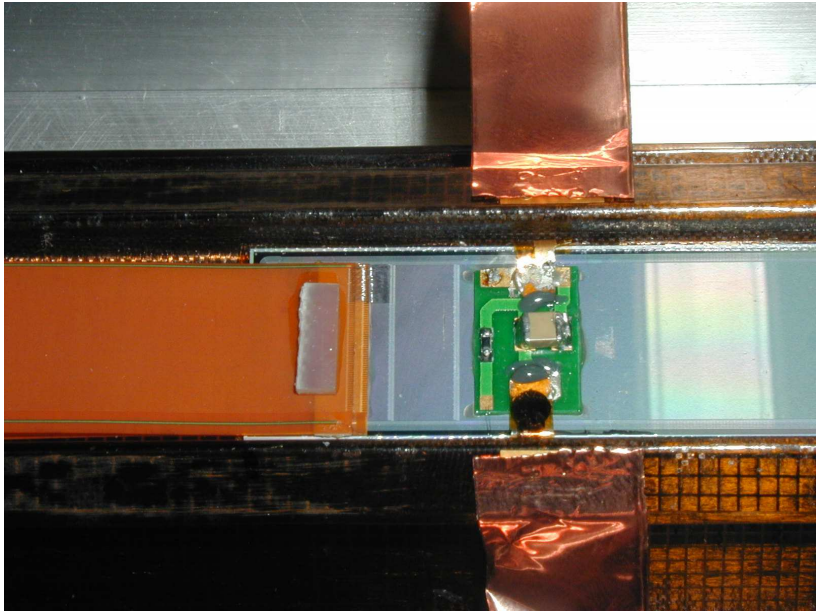}}
  \caption{
           Photograph showing details of the silicon module mounted 
           on the support structure. 
           Mounted on the sensor is the high voltage 
           filter card with the ground connection. Clearly visible are
           the ground mesh and the copper connections to the Faraday
           cage. }
  \label{fig:photo}
  \end{center}
\end{figure}

The silicon sensors are mounted on the outer shell
with the embedded grounding copper-kapton mesh using epoxy. 
The $p$-side of the sensor was grounded to the carbon fiber
support structure through the embedded copper-clad kapton 
flex circuit as described above. The rectangular ground tab on the 
`wrap-around' flex circuit was attached to the 
biasing filter card, that was glued on top of the 
sensor, and the round pad on the backside sensor lamination 
picks up the sensor bias voltage (see Fig.~\ref{fig:flex}). 
The sensor bias line was wirebonded to the ground pad on the 
bias filter card.  
The two-layer analogue cable was wirebonded to the sensor at one end
and to the ceramic hybrid with two SVX4 chips at the other end. 
The hybrid was mounted on the hybrid support structure.
A photograph of a detail of the silicon module on the support 
structure is shown in Fig.~\ref{fig:photo}. 
Although the hybrid can accommodate 
two readout chips, only the first readout chip was wirebonded. 
The first 64 channels of the readout chip were only connected to 
the analogue cable and the second 64 channels were also connected to 
the sensor. Of this second set of 64 channels, the last 30 channels
were inactive. 
The DC resistance of the 47 cm long sensor support structure 
was 12~$\Omega$ and the resistance of the ground trace on the 
analogue cable was 20~$\Omega$.  
The whole assembly was mounted inside an aluminum Faraday cage.
The Faraday cage was placed on copper-clad G10 which was 
connected to the ground of the readout system and of the HV power 
supply. The electrical connection between the copper-clad G10 
and the Faraday
cage was quite good; the DC resistance was measured to be less than
1~$\Omega$.

\section{Test Results}
\label{sec:results}

First, the test results of the L\O\  module before 
installing it on the prototype support structure are given.
This result serves as a reference for the effect of adding
the support structure.

We define the noise on the output of a given channel 
as the standard deviation $\sigma$ of the digital output 
in ADC counts for that channel: 
\begin{equation}
\sigma_a^2 \,=\, \overline{a^2} - \overline{a}^2  \ . 
\end{equation}
In practice, the quantity $\sigma_a$ is estimated by computing the 
root mean square value of the digital output of the channel under 
consideration over a large number of events (typically 1000). 
Because we are interested in the susceptibility of the system to 
external noise sources, we also define the ``differential'' noise
of a given channel as $1/{\sqrt 2}$ of the standard deviation of 
the difference between the output signal $a$ for that channel and 
the digital output signal $a'$ of a neighboring channel: 
\begin{equation}
(\sigma_{a}^{diff})^2 \,=\, \frac {\overline{(a - a')^2} \,-\, 
                                   \overline{(a - a')}^2  }
                                  {2}   
\end{equation}
Again, in practice the quantity $\sigma_a^{diff}$ is estimated 
by computing the root mean square value of $(a-a')/\sqrt{2}$ over a 
large number of events. Assuming that 
$\sigma_a = \sigma_{a^\prime}$, i.e. 
that the channel under consideration and its neighbor
have the same noise, we have
\begin{eqnarray}
(\sigma_{a}^{diff})^2 
                  & = & \frac {\sigma_a^2 + \sigma_{a^\prime}^2} 
                                 {2}   
                          \,-\,  \overline{aa'} 
                          \,+\,  \overline{a} \,\, \overline{a'} \\
                  & = & \frac {\sigma_a^2 + \sigma_{a^\prime}^2} 
                                 {2}   
                          \,-\,  \rho \, \sigma_a \, \sigma_{a^\prime}  \\
                  & = & \sigma_a^2 \, ( 1 - \rho) , 
\label{eq:noise}
\end{eqnarray}
where $\rho$ is the correlation coefficient between $a$ and $a'$, 
$\rho = cov(a, a') / ( \sigma_a \sigma_{a^\prime} ) $.
The differential noise is thus equal to the total noise if there 
is no correlation between the output signals of neighboring channels. 
A positive correlation can be due to a common component of noise 
induced by external pickup and is the object of study. 
Due to capacitive coupling between neighboring strips in the silicon
sensor a negative correlation could be present as well at the few
percent level~\cite{ref:lutz}. This effect has been neglected here. 
When the total noise of the device is equal to the differential  
noise and is the same as the value obtained on the test bench in
a Faraday cage, the performance of the system is considered to be 
optimal.

\subsection{Before Installation}

The silicon module was tested on the bench inside a Faraday cage. 
The total noise before installation on the carbon fiber support structure
was measured to be 2.7~ADC counts for channels connected 
to both the sensor and the analog cable with 
a vanishing correlation coefficient. That is, there was no 
significant difference between total and differential noise. 
The bandwidth setting of the SVX4 was such that the 10-90\% 
risetime was measured to be 35~ns and
65~ns for 10~pF and 33~pF capacitive load, respectively.
The integration time of the preamplifier was 132~ns.
The resulting risetime is fast enough to collect all charge
for the capacitive loads used in the measurements. 
All results are quoted in ADC counts. 
For the parameter settings used for our tests the gain of the 
SVX4 chip is approximately 700 electrons per ADC count. The measured noise 
of 2.7~ADC counts, or 1890~$e^-$, is in reasonable agreement 
with the expected noise, given a strip load capacitance of 
1.4~pF/cm.


\subsection{After Installation}

\begin{figure}[htbp]
  \begin{center}
  \resizebox{!}{3.5in}
               {\includegraphics{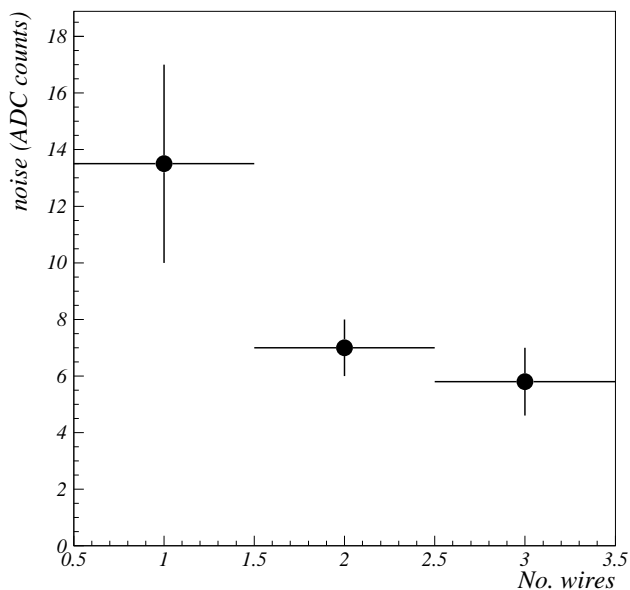}}
  \caption{
           Noise in terms of ADC counts as a function of number of
           wires connecting the hybrid ground and the Faraday cage.
          }
  \label{fig:noise_nwires}
  \end{center}
\end{figure}

The module was then installed on the support structure and the whole 
assembly was placed in a Faraday Cage.
The ground connection between the hybrid and the support structure 
was established by soldering a wire from the hybrid ground pad to 
the copper strip on the embedded copper-kapton mesh circuit, 
embedded in the support structure. 
The support structure itself was connected to the Faraday cage 
with copper tape. One end of a copper strip was taped to the 
chassis of the Faraday cage, and the other end to the gold pads 
on the kapton flex embedded in the support structure.
The measured noise level was about 14~ADC counts, 
considerably higher than the performance on the bench. 
This indicates a persisting high impedance ground connection. 
Both resistance and inductance contribute to the impedance, but
for high frequencies the inductance often dominates.
The ground connection at the hybrid end was then modified by 
adding wires connected in parallel between the hybrid and the 
support structure. 
Additional wires can be viewed as adding inductances in 
parallel, thus reducing the overall inductance. 
Fig.~\ref{fig:noise_nwires} 
shows the total noise as a function of the number of wires 
used in grounding the hybrid to the support ground. 
Because the resistance of each wire is quite small, much less than
the other connections in the ground path, the dependence of the noise
on the configuration of the wires is due to the inductance of the wires.
Shortening the wire length from 4~cm to approximately 1.5~cm suppresses 
the noise by about 10\%.

The setup was then carefully improved by enlarging the ground  
connection to the support structure and minimizing its inductance
using copper braid. 
In addition, one end of a thin strip was soldered directly to the 
ground on the hybrid and the other end taped to the Faraday 
cage using copper tape. This resulted in the same noise level as 
before the module was installed on the support structure. 

\begin{figure}[tp]
\centering 
\includegraphics[width=4.5in]{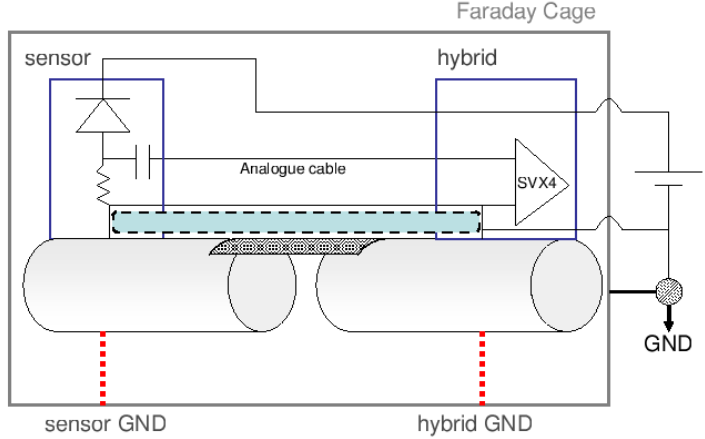}
\caption{Equivalent circuit diagram for the final test configuration 
         with additional ground connections between the two support 
         structure. }
\label{fig:equiv_ground}
\end{figure}

Next, two modifications were made to this single-point ground 
configuration. 
First, the sensor was grounded to the sensor support structure, 
which in turn was connected to the Faraday cage. This was 
done by connecting the ground on the high voltage filter 
card, mounted on top of the sensor, to the sensor support structure, 
which in turn was connected to the Faraday cage with copper tape.
One end of a copper strip was taped to the chassis of the Faraday cage,
and the other end to the gold pads on the kapton flex
embedded in the support structure.
Secondly, the grounds of the sensor support structure and the 
hybrid support structure were connected. 
One end of a copper ground mesh circuit was attached with silver 
epoxy to the gold pad on the kapton flex circuit that was co-cured 
onto the sensor support structure and the other end to the gold pad 
on the hybrid support structure (see Fig.~\ref{fig:solidmodel}). 
This minimizes the area of the ground loop. The ground return is 
now a continuous, low inductance mesh circuit in close 
proximity to the analogue cable and the hybrid. 
The equivalent circuit diagram is shown in Fig.~\ref{fig:equiv_ground}.

\begin{figure}[htbp]
  \begin{center}
  \resizebox{!}{2.8in}
               {\includegraphics{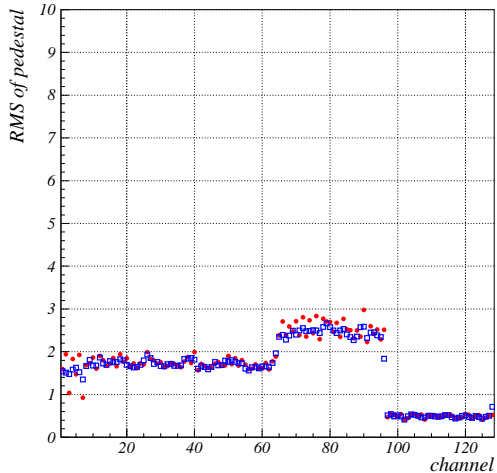}}
  \caption{Noise distribution in ADC counts versus channel number 
                after installing the silicon module onto the support 
                structure. 
                The total noise is represented by solid circles, the
               differential noise by open squares.  }
  \label{fig:noise_support}
  \end{center}
\end{figure}

The measured noise distribution is shown in Fig.~\ref{fig:noise_support}
and is identical to the behavior of the module on the bench. 
Both the total noise (solid circles) and differential noise 
(open squares) is plotted. 
An increase in noise is observed at channel 64. Recall that the first
64 channels were only connected to the analogue cable, whereas the
following 34 channels were also connected to the sensor. The increase
in noise from about 1.8 to 2.7~ADC counts is consistent with the added
capacitive load of 11.2~pF of the silicon strip. 
The last 30 channels were inactive. 
Note that, due to the double stack configuration of the analogue
cable, a small odd-even effect in the pedestal distribution persists: 
the readout channels in the bottom layer of the analogue cable 
are more susceptible to external noise and have a larger pedestal 
rms value. The assumption which lies 
at the basis of equation~\ref{eq:noise}, namely that the rms of 
neighboring channels is the same, is not completely valid in this 
situation. Odd and even channels have different proximity to the 
ground plane and thus an intrinsic different noise behavior.

\section{General Grounding Scheme}
\label{sec:gnd_scheme}

We have developed a method of integrating grounding circuits 
in a carbon fiber support structure and have demonstrated that 
embedding copper-kapton mesh circuits in the various support 
elements, with the copper covering more than 15\% of the area, 
is a flexible way of providing ground connections that
is extremely robust. 

Two grounding configurations were studied. 
First, a single-point ground at the hybrids was studied. 
Theoretically, a single-point ground provides the best 
grounding configuration. In practice, however, 
it is difficult to achieve a single-point ground
especially for systems like the close-packed L\O\ 
structure, because of the tight space constraints.
For example, the support structures of nearby detector elements 
can easily couple to L\O\ capacitively, resulting in a breakdown of 
the single-point grounding scheme. Because of the inherent 
uncertainties when installing a detector as part of a complete 
system and operating all elements of the system concurrently, 
it is most prudent to implement a multi-point ground. 
The challenge then is to achieve 
the same noise performance with a multi-point ground. 
Our test results show that the same noise performance 
as with the single-point ground, with the sensors floating, 
is obtained. 
Since there is no improvement using a single-point ground
and because such a system is difficult to implement in practice, 
the safer approach of carefully grounding everything together and 
minimizing any potential ground loops was adopted. 
Our test results indicate that in this configuration two key 
requirements for low noise performance are to be respected: 
the ground loop 
must be minimized, and there must be a low impedance ground 
connection between the active detector and the front-end 
readout electronics. 

For the L\O\ silicon detector at small radius being proposed for the 
D\O\ experiment, these general grounding principles have been 
implemented and excellent results were obtained. 
A copper-kapton mesh circuit was embedded in the sensor 
support structure with the kapton covering the total surface area
and the copper having an area coverage of 30\%.  
With this ground mesh embedded in the 
surface of the support structure it was straightforward to 
make low inductance ground connections to the sensor 
using small copper-kapton jumper cables attached to 
precisely positioned gold plated pads in the ground mesh using 
silver epoxy.  

To minimize the inductance of the ground path, 
a uniform continuous ground plane between the
sensor and hybrid is needed. 
Another copper-kapton mesh circuit, with the same specifications, 
was embedded in the hybrid support structure. 
Small copper-kapton jumper cables attached to gold-plated pads 
at the end of either support structure provided the electrical 
connection between them. 
All hybrids and all sensors are thus tied to one uniform 
ground mesh.
Grounding the hybrid was a bit more difficult due to space constraints. 
Based on the rule of low inductance connections, the shortest path from
the hybrid to the inner support cylinder 
is needed. This has led to a design where the hybrid has a ground
strip at the bottom which makes direct contact with the support 
structure.


\section{Carbon Fiber Conductivity}

In this paper a pragmatic approach has been taken in the design 
of electrically robust detectors using carbon fiber. 
It is still an intriguing question why carbon 
fiber is so highly conductive at high frequencies. 
We speculate that this is due to the large skin depth. 
Carbon fiber consists of strands of conductor embedded in a 
non-conductive epoxy. This configuration nearly eliminates eddy 
currents, which otherwise would reduce conductivity limiting it 
to the surface. It is similar to the use of plates in conventional 
motors or transformers to get rid of eddy currents.  Without eddy 
currents, the area available for conduction in carbon fiber is much 
thicker than for a similar amount of copper. This conjecture, 
however, needs to be experimentally verified.

\section{Conclusion}
\label{sec:conclusion}

A low inductance grounding scheme with the smallest possible 
ground loop is essential to achieve low noise performance. 
This implies the use of short and wide electrical
connections, and a careful mechanical design to minimize or 
eliminate ground loops.
We have shown that for frequencies relevant for operation of 
silicon detectors in the Tevatron environment, carbon 
fiber is a good conductor and cannot be distinguished from 
copper. We have demonstrated that the use of copper-kapton flexible 
mesh circuits, with the copper covering more than 15\% of the 
surface, provides adequate electrical coupling to carbon fiber 
support elements. Co-curing these circuits with the carbon fiber 
during the lay-up process provides an electrically integrated 
support, with great flexibility to adapt to the specific 
electrical requirements for that application.  
In a future paper the results of extensive tests on a small 
radius silicon detector, following the design rules outlined 
above, will be presented.

\section{Acknowledgments}

We wish to thank Kurt Krempetz, 
Mike Hrycyk and Frank Lehner for valuable advice, 
Bert Gonzalez for the assembly 
of the modules and all our D\O\ 
silicon colleagues for helpful discussions.


\end{document}